\documentclass[preprint,preprintnumbers,amsmath,amssymb]{revtex4}
\usepackage{amssymb}
\usepackage{txfonts}

\usepackage{graphicx}
\usepackage{dcolumn}
\usepackage{bm}

\begin{document}

\title{Proof-of-randomness protocol for blockchain consensus:\\ a case of Macau algorithms}

\author{Wen-Zhuo Zhang}
\author{Victor Kai}
\affiliation{GPD Labs, QUAKEY Ltd.}

\begin{abstract}
A proof-of-randomness (PoR) protocol is presented as a fair and low energy-cost consensus mechanism for blockchains. Each network node of a blockchain may use a true random number generator (TRNG) and hash algorism to fulfil the PoR protocol. In this paper, we give the consensus mechanism of the PoR protocol, and classify it into a new kind of randomized algorithms called Macau. The PoR protocol could generate a blockchain without any competition of computing power or stake of cryptocurrency. Besides, we give some advantages of integrating quantum random number generator (QRNG) chips into hardware wallets, and also discuss the way to cooperate with quantum key distribution (QKD) technology.
\end{abstract}

\maketitle

\section{introduction}

Blockchian is a decentralized data recording technology of a network briefly for on-line trading. All network nodes are distributed in space, while blocks are distributed in time-order with a chain connection. The current block of a blockchain usually records its generation timestamp, the nonce of the node who generates the current block, the hash value of the last block's data, and all transactions in the network after last block. Consensus algorithms play the central role of generating blocks. The first blockchain, bitcoin\cite{bitcoin}, use proof-of-work (PoW) consensus to select network nodes for recoding a new block. The PoW requires computers to calculate a hash value with very small probability. For example, a 256-bit hash value with all zeros in highest $\sim70$ bits. Such method consumes huge computing power which means a waste of massive electricity energy. Recently, Ethereum \cite{ethereum} replace its PoW by proof-of-stake (PoS) method to save more than $99\%$ power, however, the fairness of the PoS is questioned.

In this paper, we present a fair and low-energy cost consensus algorithm called proof-of-randomness (PoR) protocol. The PoR requires true random number generator (TRNG) hardware on each network node. The fairness comes form the physically true random number, and the main power consumption is from a TRNG chip which is usually less than 10mW on each node. Since PoR depends on multiple random number inputs to get final output, it is not include in either Monte Carlo or Las Vegas randomized algorithms. We define this new kind of randomized algorithms as \emph{Macau}, and discuss the general features of Macau randomized algorithms.

\section{generation and testing of true random numbers}

Before running PoR protocol, every network node should acquire an array of binary true random numbers form a TRNG hardware. For stationary computers, such as servers or desktop PCs, a big TRNG device or a PCI card with TRNG chips could work. However, for mobile computers such as laptops, smart phones or tablet PCs, a mobile USB device with quantum random number generator (QRNG) chips is a good choice. 

Quantum randomness is the fundamental true randomness of the universe. Einstein once do not believe quantum randomness and said "God never play dice", however, quantum randomness has been proved by all quantum mechanics experiments and indeed \emph{God plays dice.} Since QRNGs perform much better than classical TRNGs in complex environment, and also has much lower energy consumption and much smaller size than classical TRNG chips \cite{IDQ-whitepaper}, a small-size or mobile hardware with a QRNG chip is a suitable solution for blockchain network nodes.

We can test the quality of random numbers by embedding testsuits into blockchain softwares, e.g., NIST SP 800-22 randomness testsuit\cite{NIST-test}. This step is necessary to forbid any pseudo random number generator (PRNG) since pseudo random number are indeed determined and not fair enough for PoR. Typically, it requires at least 1 MB binary random numbers to finish the randomness test. When a node's QRNG (or TRNG) chip has output 1 MB random number data and passed the test, blockchain softwares could split the 1 MB data into 1024 pieces, for example, with each piece's length at 1Mb, and select out one 1Mb length random numbers. We name the selected random numbers as $R_{n}(m)$, with the block number $m$ and the network node number $n$.

\section{proof-of-randomness protocol}

When a network node finishes its true random number test, the selected 1 Mb random number $R_{n}(m)$'s hash value is calculated by SHA-1024 and get a 1024-bit hash value $H_{n}(m)$. The hash value is then sent to other nodes together with the node's digital signature.

In the case that the network is relatively small, the synchronizing of random number data among all $N$ online nodes could be within several minutes. Then each node receives every other online node's 1Mb random number and every node acquires the same summation of 1 Mb random numbers, which is
\begin{equation}\label{sum}
S_H(m)=\sum_{n=1}^{N}H_{n}(m).
\end{equation}
This sum is independent to the order of receiving $R_{n}(m)$ on every node due to commutative law. Then every node calculate the hash value of $S_{H}(m)$ by SHA-1024 and broadcast the result $H(m)$ to all other nodes together with its digital signature. Here we call $H(m)$ the total hash value of block $m$ in order to distinguish the individual hash value $H_{n}(m)$ from every node. Since $H(m)$ depends on all 1 Mb randome number $R_{n}(m)$ form every node, it is truly unpredictable, and the absolute difference value $D_{n}(m)=|H(m)-H_{n}(m)|$ is also unpredictable. It prevents the attacks of using controllable pseudo random numbers form any node. Then a consensus of choosing the node with minimal or maximal $D_{n}(m)$ as the $mth$ block's owner or recorder is build.

\begin{figure}
\includegraphics[width=150mm]{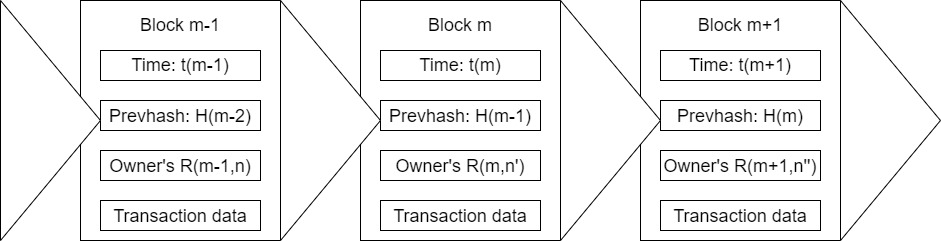}
\caption{In proof-of-randomness (PoR), a block records the owner's 1Mb true random number instead of a \emph{nonce} in PoW.}\label{blocks}
\end{figure}

In the case that the network is relatively big, synchronizing of data among all nodes is not feasible. It would probably happen in large public blockchains. Here we should not calculate the total hash value $H(m)$ anymore since different nodes may get different $H(m)$. Instead, every node could calculate the hash value of $H_n(m)+H(m-1)$ by SHA-1024 and broadcast the result $H'_{n}(m)$, where $H(m-1)$ is the total hash value of the last block. Then every node calculate $D_{n}(m)=|H'_{n}(m)-H_{n}(m)|$ and broadcast $D_{n}(m)$ together with its digital signature. A consensus of choosing the node with minimal or maximal $D_{n}(m)$ as the $mth$ block's owner still works.

A kind of attack may happen in relatively big network, which is to try multiple generation of random numbers and select a much closer $R_{n}(m)$ to the target minimal or maximal $D_{n}(m)$. A node with more QRNG chips or with expensive TRNG devices could do the attack. In order to prevent such kind of drawback to PoW (with a competition of random number generating power instead of computing power), a node $n$ should broadcast its time-stamp of finishing the true random number test, $t_{n}(m)$, and its time-stamp of finishing $D_{n}(m)$, which is $t_{n}'(m)$. A consensus of choosing the node with minimal $D_{n}(m)\times\Delta t(m,n)$ as the $mth$ block's owner could be reasonable, where $\Delta t_{n}(m)=t_{n}(m)-t_{n}'(m)$. If a node try to generate random numbers by multiple times, $\Delta t(m,n)$S would increase and reduce its possibility of get minimal $D_{n}(m)\Delta t_{n}(m)$. Besides, $\Delta t_{n}(m)$ should also set a minimal value in order not to be attacked by forging timestamps.

\section{Randomized algorithm: Macau}

There are two kinds of widely used randomized algorithms. One is Monte Carlo, which use limited steps of random inputs to obtain a output with a high probability to be the predetermined target value. Monte Carlo is the main randomized algorithm for statistics and simulations. The other is Las Vegas, which use unlimited steps of random inputs to obtain a output that must be the predetermined target value. PoW is a typical example of Las Vegas algorithms, with a hash value with $\sim70$ zeros in high bits as the predetermined target, and the number of random attempts is not limited until such hash value is calculated out in a network node.

PoR is a randomized algorithm with true random number inputs, while the output depends on the hash value of all inputs. Since the output is not predetermined, PoR can not be classified into either Monte Carlo or Las Vegas algorithms. Here we define a new kind of randomized algorithms called Macau to classify the PoR. Compared to Monte Carlo and Las Vegas, Macau algorithms use limited steps of random inputs to obtain a output that is not predetermined before any inputs.

\begin{figure}
\includegraphics[width=120mm]{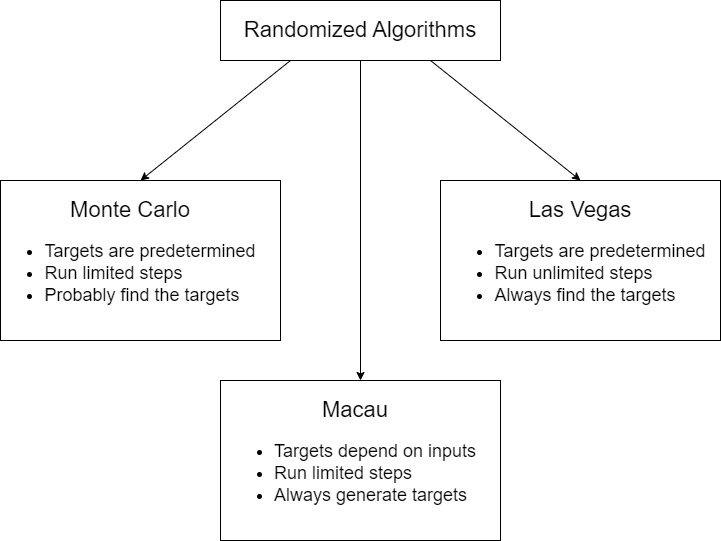}
\caption{Brief differences between Macau algorithm and Monte Carlo algorithm, as well as Las Vegas algorithm.}\label{randomized}
\end{figure}

We give a general description of Macau algorithms by Eq.\ref{Macau}
\begin{equation}\label{Macau}
T=F[\sum_{n=1}^{N}R(n)],
\end{equation}
where $T$ is the target value, $R(n)$ is the $nth$ random number input, and $F[x]$ is a function of the variable $x$ which is the sum of all input random numbers. In PoR, $F[x]$ is just the SHA function of the sum of input random number's hash values. Since both addition and multiplication have commutative law, the function $F[x]$ could be the product of all input random numbers too, which is
\begin{equation}\label{Macao}
T=F[\prod_{n=1}^{N}R(n)].
\end{equation}

Macau algorithms are different to Monte Carlo and Las Vegas, since both of them have predetermined target values $T_p$. The equation of either Monte Carlo or Las Vegas is like $T(m)=F[R(m)]$, where $R(m)$ is the $mth$ attempt of random number input, and $T(m)$ is the result of $mth$ attempt by function $F$. Monte Carlo algorithms limit the attempt times, where a $T(m)$ equals to $T_p$ could probably be found. Las Vegas algorithms never limit attempt times in order to find $T(m)=T_p$.

\section{other advantages of using QRNGs in blockchains}

Blockchains adopt elliptic curve cryptograph (ECC) as the main encryption and decryption method. The ellipse curve equation is
\begin{equation}\label{EC}
y^2=ax^3+bx+c,
\end{equation}
while most of current blockchains use a simplified elliptic curve equation, $y^2=x^3+7$.

There are three kind of random numbers in ECC. The first one is a base-point on the elliptic curve, $G$, which starts a half-line to cross the elliptic curve. The second one is the private key, $k$, which is the reflection times of the half-line on the elliptic curve. The public key $K$ is the final point of the half-line on the elliptic curve determined by $G$ and $k$, and expressed by a "multiplication" defined on the curve, which is $K=G*k$. Since it is extremely difficult to calculate $k$ with public $G$ and $K$ values by a classical computer, the ECC is secure to all classical computers now.

The third random number is $R$, which encrypt the data $D$ together with the public key $K$. The encryption equations are
\begin{eqnarray}\label{en}
E_1=K*R+D \\
E_2=G*R,
\end{eqnarray}
while the decryption equation with private key $k$ is
\begin{equation}\label{de}
E_1-E_2*k=K*R+D-G*k*R=D.
\end{equation}
Here we can see when $R$ is not safe, e.g. a pseudo random number that could be attacked by stealing its short seed, an attacker could obtain $D$ with $D=E_1-K*R$ since both $K$ and $E_1$ are public.

When QRNGs are evolved in network nodes, both $k$ and $R$ can use quantum random numbers to prevent any crack by computers. It makes $E_1$ much safer to travel over the network. Therefore, a hardware wallet with a QRNG could upgrade ECC's security, which is also an upgrade of blockchain's security.

Besides, QRNG chips in hardware wallets could provide quantum keys to encrypt their private keys in order to recover them. For current hardware wallets, when a wallet is lost, the private key is lost too. Then all the cryptocurrency bounding to this private key would be lost forever. When a hardware wallet use a quantum key $q$ from a QRNG chip to encrypt its private key $k$, the encrypted data $E$ could be sent out the wallet and stored in a disk or on cloud serve. While the quantum key could be stored in an off-line memory device which is apart from $E$. If a hardware wallet is lost, one can recover its private key by decrypting $E$ with $q$. This is another advantage of hardware wallets with QRNG chips.

\section{in cooperation with quantum key distribution (QKD)}

Since ECC is vulnerable to general quantum computers, blockchains should upgrade their cryptograph to keep the future security when a general quantum computer with millions of qubits comes to reality. Post-quantum cryptograph (PQC) is a natural solution which costs a general quantum computer much longer time to break than breaking RSA and ECC. However, it can not guarantee that a future quantum algorithm which breaks the PQC in a very short time never exist.

Quantum key distribution (QKD) is the ultimate solution of preventing the attack form quantum computers \cite{QKD}. In QKD, all the keys are quantum random numbers, and any interception of key distribution would break the key generation. Therefore, QKD with one-time encryption is unconditional secure to any algorithm. If every node of a network have a server together with a QKD terminal, one can use one-time keys form QKD network to do peer-to-peer authentication between any two nodes.

However, QKD networks rely on optical fibers or even satellites like Mucius which restrict QKD terminals in cloud servers. QKD terminals can not be used in mobile networks either since there is no single-photon technology in radio-frequency or microwave band. Here we could use QRNG chips in a mobile node to do an off-line local distribution of quantum keys to any server node with QKD terminals. Then an on-line key update could work with current keys encrypting next keys (KEK) for communication. This method could work in all mobile networks since it is indeed a classical communication encrypted by quantum keys. With this method being applied on mobile networks, and QKD being applied on fiber networks, a full quantum-safe network is feasible.

\section{conclusions}

In this paper, we present a proof-of-randomness (PoR) protocol. Such protocol could give a fair and low energy-cost consensus mechanism to blockchains with true random number generators (TRNG) and hash algorithm. Since low-cost quantum number generators (QRNG) make the true random number resource more accessible to common users, the PoR protocol could popularize the blockchain technology, especially public blockchains for Web3 applications. We classify PoR to a new kind of randomized algorithms called Macau algorithms in order to distinguish PoR form widely-used Monte Carlo and Las Vegas algorithms.

Besides, a hardware wallet with a QRNG chip could not only give out quantum random numbers to do the PoR protocol, but also make elliptic curve cryptograph (ECC) much safer by replacing all pseudo random numbers with quantum random numbers. Further more, a quantum random number could be a quantum key to encrypt the private key in the wallet, then a backup of the encrypted private key outside a hardware wallet is possible.

Finally, QRNG chips could cooperate with quantum key distribution (QKD) in order to prevent any attack form general quantum computers to blockchains in the future.

\section{acknowledgement}

As a MiraclePlus alumni company (F22), we thank MP for its accelerator program and founders' community.


\begin{thebibliography}{10}

\bibitem{bitcoin}
https://nakamotoinstitute.org/bitcoin/

\bibitem{ethereum}
https://ethereum.org/en/whitepaper/

\bibitem{IDQ-whitepaper}
https://www.idquantique.com/random-number-generation/overview/

\bibitem{NIST-test}
https://nvlpubs.nist.gov/nistpubs/Legacy/SP/nistspecialpublication800-22r1a.pdf; 

\bibitem{QKD}
V. Scarani, et. al., Rev. Mod. Phys. 81, 1301 (2009); F. Xu, X. Ma, Q. Zhang, H. Lo, and J. Pan, Rev. Mod. Phys. 92, 025002 (2020)

\end{thebibliography}
\end{document}